\documentclass[runningheads]{llncs}

\usepackage[T1]{fontenc}
\usepackage{graphicx}
\usepackage{wrapfig}
\usepackage{amsmath}
\usepackage{amssymb}
\usepackage{booktabs}
\usepackage{array}
\usepackage{tabularx}
\usepackage{arydshln}
\usepackage{xcolor}
\usepackage{listings}
\usepackage{float}
\usepackage[hidelinks]{hyperref}

\graphicspath{{figures/}{./}}

\begin{document}

\title{Who Changed the AI Agent's Rules? Measuring How AI Coding Agents Escalate Authority in CI/CD Pipelines}
\titlerunning{Measuring How AI Coding Agents Escalate Authority in CI/CD}

\author{Miroslaw Staron}
\authorrunning{M. Staron}

\institute{Department of Computer Science and Engineering,\\
University of Gothenburg and Chalmers University of Technology,\\
41296 Gothenburg, Sweden\\
\email{miroslaw.staron@gu.se}}

\maketitle

\begin{abstract}
AI coding agents can modify CI/CD control-plane structures, including workflows and permission settings, thereby extending their authority beyond routine code changes. Because such modifications can affect an entire development pipeline, this study examines the trade-off between execution cost and governance risk in agentic CI/CD. We introduce a joint penalty model, $P(M)$, that combines these two dimensions and evaluate it on all 932,791 agent-authored pull requests in the AIDev dataset, covering five autonomous coding agents. Each pull request is automatically classified by authority scope---data-plane or control-plane---and assigned a penalty score.

We found that control-plane actions have a 2.3$\times$ higher mean joint penalty than data-plane actions (0.498 vs.\ 0.218) and a 3.9$\times$ higher mean governance risk (0.530 vs.\ 0.135), which means that they indeed increase the risk. All 775 pull requests in the extreme-risk category, defined as $P(M) > 0.8$, modify the control plane. Policy-violation flags occur in 46.6\% of all control-plane pull requests and in 63.6\% of merged control-plane pull requests. We have also found an indication that governance risk differs substantially across agent implementations, ranging from 0.166 for Cursor to 0.328 for Copilot in mean joint penalty. These differences are not explained by execution cost or merge rate, suggesting that authority policies should be calibrated for individual agents. Manual assessment of 34 pull requests yielded a Cohen's $\kappa$ of 0.72 for the authority-scope classifier.

The results show that a single compound metric can identify and measure changes to CI/CD governance structures at scale. The proposed framework supports detecting high-risk pull requests and comparing governance behavior across agent implementations, providing a basis for authority tuning and pipeline governance in agentic software development.

\keywords{CI/CD \and Agentic systems \and Governance risk \and Authority escalation \and Autonomic computing}
\end{abstract}

\section{Introduction}
\label{sec:introduction}
\noindent
Autonomous coding agents -- including OpenAI Codex, GitHub Copilot, Devin, Cursor, and Claude Code -- are moving beyond code completion and taking a more active role in the software development lifecycle. They generate code, write tests, respond to review comments, and author pull requests (PRs). In some configurations, they can also merge their own changes into production branches. The scale of this activity is already considerable as the AIDev dataset contains 932,791 agent-authored PRs from more than 116,000 repositories~\cite{aidev,aidev-teammates}.

Both human-authored and agentic contributions pass through the same CI/CD infrastructure. That infrastructure builds, tests, reviews, approves, and deploys software regardless of the author. Because the infrastructure can be configured and altered programmatically, agents gain more authority in contemporary software development. As agents become more involved in both development and CI/CD configuration, it becomes important to distinguish changes that operate within existing rules from changes that alter the rules themselves. Current software-development metrics do not make this distinction explicit.

Most agent-authored PRs affect the \textit{data plane}\footnote{We can define the data plane to be the standard software development of new product features, improvements, etc. It is an analogy to the data plane in telecommunication systems: voice and data transferred via the network.}~\cite{assistance-to-agency}: application code, tests, documentation, or dependencies. Such changes are evaluated within existing pipeline constraints and can normally be reviewed, rejected, or reverted without altering the governance mechanism of CI/CD itself. A smaller but more consequential class of changes affects the \textit{control plane}\footnote{We define the control plane similar to the signalling of the telecommunication networks: special kinds of packets that (re-)configure the network. By analogy the control plane is defined as changes that alter the CI/CD pipeline, e.g., (re-)configure approvals.}: CI/CD workflows, approval gates, permission settings, secrets, deployment policies, and related governance structures.

For example, an agent that edits \texttt{.github/workflows/} is not modifying software behavior, but it is changing the mechanism that determines which future changes may proceed. We call this \textit{authority escalation}: an agent modifies structures that define or constrain its own authority. The consequences may extend beyond the PR that introduced the change because subsequent contributions, whether authored by humans or agents, are evaluated under the revised rules.

Traditional metrics are poorly suited to detecting this situation. Measures such as merge latency, CI success rate, deployment frequency, or code-quality scores describe how changes perform within a pipeline. They do not capture whether a change affects the pipeline's governance. As a consequence, a PR that removes an approval gate may pass every automated check and merge more quickly than a routine bug fix, even though its organizational consequences are substantially greater.

To capture this distinction, we introduce a joint penalty model metric:

\begin{equation}
P(M) = \omega_1 \cdot C_{\text{exec}}(M)
     + \omega_2 \cdot R_{\text{gov}}(M),
\end{equation}

where $M$ denotes a merge candidate PR. The model evaluates each PR along two dimensions: execution cost and governance risk. The first captures the effort needed to (potentially) revert the changes, and the latter captures the potential effect of the change. The \textit{Execution cost}, $C_{\text{exec}}$, represents the computational and organizational effort associated with the change, using normalized task complexity and review latency. The \textit{Governance risk}, $R_{\text{gov}}$, captures three observable outcomes: human rejection, policy violation, and post-merge regression. It also includes a structural coefficient, $\kappa = 0.25$, for control-plane changes, reflecting the additional risk associated with modifying governance structures.

The default weights, $\omega_1 = 0.4$ and $\omega_2 = 0.6$, place greater emphasis on governance risk. Organizations can adjust these weights to reflect their priorities. Increasing $\omega_1$, for example, gives greater importance to execution efficiency, whereas increasing $\omega_2$ makes the model more sensitive to governance concerns. The model can therefore compare or simulate alternative CI/CD governance policies.

We investigate the following research question:

\begin{quote}
\textbf{RQ:} How can the trade-off between execution cost and governance risk in agentic CI/CD pipelines be quantified, and how does governance risk vary across authority scopes (data-plane vs.\ control-plane) and agent implementations (Claude Code vs.\ Copilot vs.\ Devin vs.\ Cursor vs.\ Codex)?
\end{quote}

\noindent
The research question is divided into two sub-questions:

\begin{enumerate}
    \item \textbf{RQ1:} Do control-plane PRs systematically carry a higher joint penalty than data-plane PRs?
    \item \textbf{RQ2:} Does governance risk differ across autonomous coding agents, and can the framework distinguish between their risk profiles?
\end{enumerate}

We address these questions through a computational study of the AIDev dataset~\cite{aidev,aidev-teammates}, which contains 932,791 agent-authored PRs from 116,211 repositories. We compute $C_{\text{exec}}$, $R_{\text{gov}}$, and $P(M)$ for every PR. We classify each PR as either data-plane or control-plane by matching its title and description against a curated set of control-plane indicators. We evaluated the classifier through an independent manual assessment of a stratified sample of 34 PRs. The automatic and manual classifications achieved 94.1\% raw agreement and a Cohen's $\kappa$ of 0.72. We then stratified the resulting measurements by authority scope and agent, producing ten groups for comparison.

The analysis shows that for this dataset, the control-plane PRs have a 2.3$\times$ higher mean penalty and a 3.9$\times$ higher mean governance risk than data-plane PRs. We have also found that all 775 PRs with $P(M) > 0.8$ modify the control plane. Among merged control-plane PRs, 63.6\% attempt to change governance. Governance risk also varies substantially among agents and is only weakly related to execution cost. 

This paper makes three contributions:

\begin{enumerate}
    \item \textbf{C1---A measurement model for authority escalation.}
    We define the joint penalty model metric $P(M)$, which combines execution cost and governance risk for agent-authored changes. It can be computed from standard PR metadata in linear time and is therefore suitable for use as a pipeline check.

    \item \textbf{C2---An empirical analysis of agent governance behavior.}
    We apply the model to all 932,791 PRs in the AIDev dataset and compare five autonomous coding agents. We evaluate the authority-scope classifier through a manual assessment, with 94.1\% agreement and Cohen's $\kappa = 0.72$. The resulting scores provide empirical baselines for authority-related behavior in agent-authored PRs.

    \item \textbf{C3---Practical guidance for CI/CD governance.}
    Actionable recommendations: classifying the authority scope of every agent-authored PR, requiring human approval for control-plane changes, and monitoring the distribution of $R_{\text{gov}}$ to distinguish frequent control-plane activity from risky control-plane activity.
\end{enumerate}

The remainder of the paper is organized as follows. The next subsection introduces a motivating example.  Section~\ref{sec:related} reviews related work. Section~\ref{sec:theory} presents the model's theoretical foundations, and Section~\ref{sec:methodology} describes the dataset, classification procedure, and metric construction. Section~\ref{sec:results} reports the results, and Section~\ref{sec:conclusion} concludes the paper and outlines directions for further work.






\section{Related Work}
\label{sec:related}

This work draws on three areas of research: studies introducing and using the AIDev dataset, empirical research on agent-authored software changes, and literature on authority and governance in agentic CI/CD.

The empirical basis of the study is the AIDev dataset~\cite{aidev,aidev-teammates}. It contains 932,791 PRs authored by five coding agents---OpenAI Codex, Devin, GitHub Copilot, Cursor, and Claude Code---across 116,211 repositories and involving more than 72,000 human developers. The foundational studies examine agent productivity, acceptance, and review latency, and describe autonomous coding agents as ``AI teammates'' in Software Engineering 3.0~\cite{aidev-teammates}. This work extends that line of research from code contribution to release governance.

More than 60 studies accepted in the MSR 2026 Mining Challenge use AIDev. Several examine phenomena related to individual components of the $P(M)$ framework, although they do not measure authority escalation directly. Research on agent context files~\cite{agent-readmes} reports that explicit authority boundaries are uncommon in developer guidance. The proposed model complements this work by inferring authority-related behavior from PR records.

Studies of tangled commits~\cite{agentic-refactoring} and inconsistencies between PR descriptions and code changes~\cite{message-consistency} show that agent intent is not always clearly represented in PR metadata. This limitation is relevant to our classifier, which relies on PR titles and descriptions. Other work examines test contributions~\cite{test-code}, review effort~\cite{review-effort}, and security review~\cite{security-agentic}. These topics correspond to the regression, review-latency, and control-plane components of our model.

Research on code optimization~\cite{code-optimization}, dependency risk~\cite{security-risks-deps}, and code survival~\cite{will-it-survive} likewise identifies weaknesses in validation and long-term stability that relate to $R_{\text{regress}}$ and $R_{\text{policy}}$. Additional AIDev studies address library use~\cite{library-usage}, energy consumption~\cite{energy-concerns}, issue readiness~\cite{issue-ready}, and review themes~\cite{review-themes}.

The conceptual framing follows work on authority transfer in agentic systems~\cite{assistance-to-agency}. We also draw on an industrial account of failure costs across CI/CD milestones~\cite{cicd-stages}, which identifies code merge as a key point of risk control. Automated repair systems such as Repairnator, RepairAgent, and ChatRepair typically operate within a bounded data-plane role under human authority. Similarly, the MAPE-K tradition in autonomic computing~\cite{mape-k} assumes a control loop whose authority the system designer defines; it does not directly address agents that can modify the feedback and governance mechanisms themselves.

Existing work therefore either measures properties of agent-authored contributions or argues that authority is an important design variable. To our knowledge, it provides no empirical measure of authority escalation in CI/CD. The $P(M)$ framework is intended to address this gap.

\section{Theoretical Foundations}
\label{sec:theory}

A central design problem in agentic CI/CD is the allocation of authority: which decisions an agent may make, under which constraints, and with what form of human recourse~\cite{assistance-to-agency}. We distinguish between two forms of authority.

\textit{Data-plane authority} concerns localized actions performed within an established governance structure, such as generating patches, adding tests, or rerunning a failed build. \textit{Control-plane authority} concerns changes to that governance structure, including pipeline configuration, deployment policies, permissions, and approval gates. This distinction is important because: 
\begin{itemize}
    \item Agent autonomy is normally expected to remain constrained. Control-plane changes that are merged without meaningful human review may therefore indicate that the effective authority granted to the agent exceeds the intended authority model.
    \item A coding agent's safety depends partly on the governance mechanisms around it (sometimes referred to as "guardrails"). Approval gates, permissions, tests, and deployment policies compensate for the fact that the agent itself cannot guarantee safe behavior. The governance layer is therefore a critical point of trust. Weakening it may affect not only the current change but also every later change evaluated under the modified rules.
    \item Standard CI/CD metrics were designed primarily for deterministic automation. Measures such as build duration, deployment frequency, and failure rate describe operational performance but do not represent decision quality, compliance with authority boundaries, or changes to governance structures.
\end{itemize}


\subsection{The $P(M)$ Framework}
\label{sec:pm-framework}

We operationalize the preceding concepts through the $P(M)$ metric. The governance-risk component includes four elements. $R_{\text{policy}}$ represents policy adherence, $R_{\text{override}}$ represents human rejection, and $R_{\text{regress}}$ represents evidence of instability. The term
$\kappa \cdot 1*[\text{control-plane}]$
assigns an additional structural risk to control-plane changes because they alter the conditions under which subsequent builds and deployments are governed.

\begin{figure}[!htb]
\centering
\includegraphics[width=\textwidth]{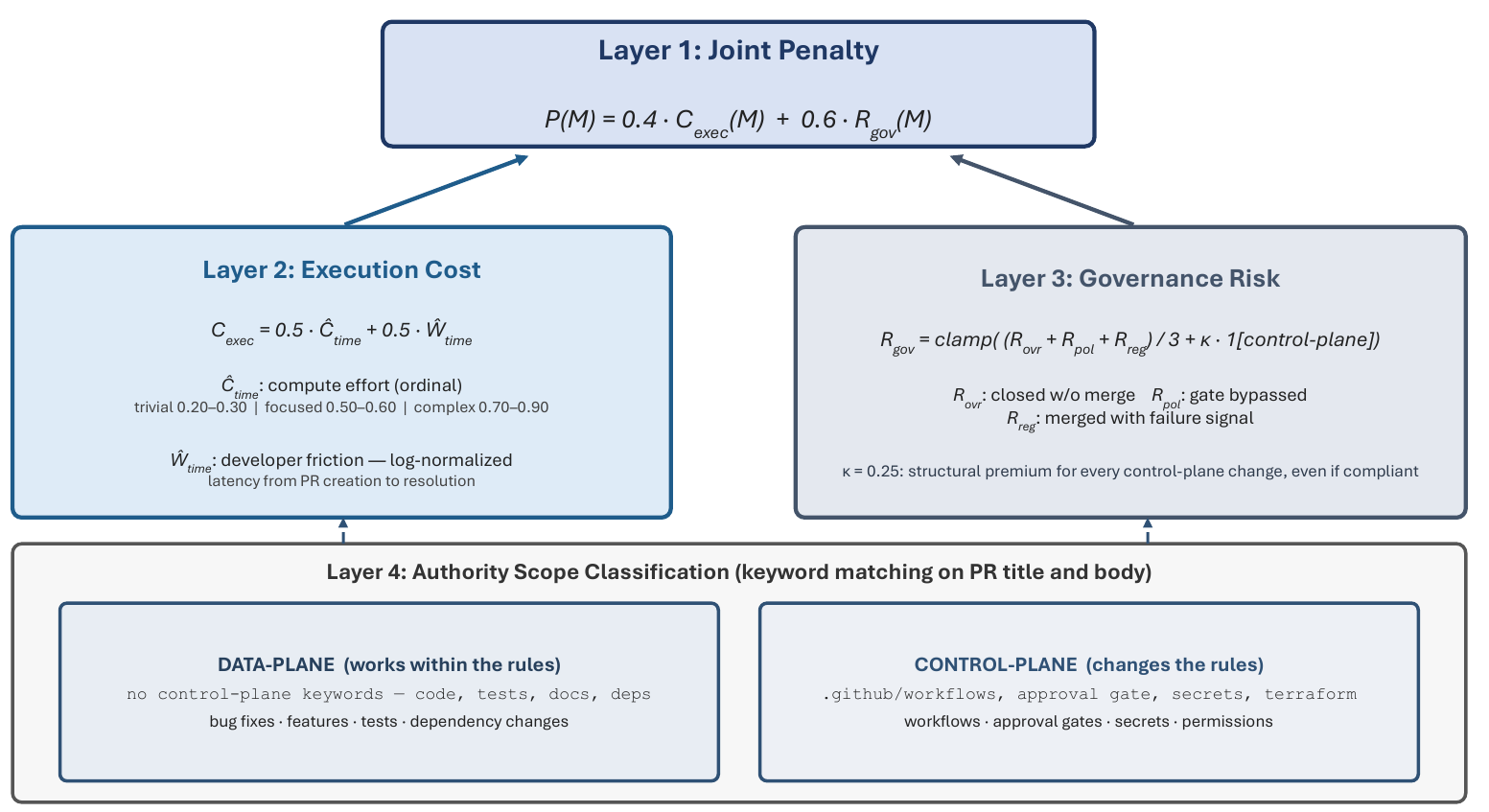}
\caption{The four-layer $P(M)$ model, comprising the joint penalty, execution-cost components, governance indicators, and authority-scope classification.}
\label{fig:pm-approach}
\end{figure}

Figure~\ref{fig:pm-approach} summarizes the four layers of the model.

\paragraph{Layer 1: Joint Penalty}: $P(M) = 0.4 \cdot C_{\text{exec}}(M)
     + 0.6 \cdot R_{\text{gov}}(M).$

\bigskip
\noindent
The weighted sum balances execution cost against governance risk. The default weights give greater importance to governance because a change to the pipeline's governing rules may have consequences beyond the immediate PR. The weights can be adjusted for different organizational settings.

\paragraph{Layer 2: Execution Cost}: $C_{\text{exec}}
=
0.5 \cdot \hat{C}_{\text{time}}
+
0.5 \cdot \hat{W}_{\text{time}}.$

\bigskip
\noindent
\textbf{Compute-time estimate}, $\hat{C}_{\text{time}}$, is an ordinal approximation of task effort. PRs are assigned to one of three categories: trivial (0.20--0.30), focused (0.50--0.60), or complex (0.70--0.90). The measure represents the estimated effort required to produce the change. It can be approximated by such measures as the engineering hours from JIRA, time from start to end of the commit. 

\textbf{Developer friction}, $\hat{W}_{\text{time}}$, is the log-normalized interval between PR creation and resolution, where resolution is either a merge or closure. Longer intervals indicate that the PR required more review or coordination (i.e., issues merging it, such as conflicts with existing code)\footnote{PRs that are open at the time of analysis have no observed resolution time and therefore receive $\hat{C}_{\text{time}} = 0$ and $\hat{W}_{\text{time}} = 0$. This ensures that the analysis is not polluted by open PRs, which is essentially the same as filtering them from the dataset.}

\paragraph{Layer 3: Governance Risk}

\begin{equation}
R_{\text{gov}}
=
\left(
\frac{
R_{\text{override}}
+
R_{\text{policy}}
+
R_{\text{regress}}
}{3}
+
\kappa \cdot 1*[\text{control-plane}],
\right).
\end{equation}
\noindent
The model uses three binary indicators:

\begin{itemize}
    \item $R_{\text{override}} = 1$ when a PR is closed without being merged, representing human rejection;
    \item $R_{\text{policy}} = 1$ when a control-plane PR is merged without an observable authorization gate;
    \item $R_{\text{regress}} = 1$ when a PR is merged despite failure signals or introduces evidence of regression.
\end{itemize}

The indicators apply to different lifecycle outcomes. $R_{\text{override}}$ applies only to unmerged PRs, whereas $R_{\text{policy}}$ and $R_{\text{regress}}$ apply only to merged PRs. Consequently, a data-plane PR can have at most one active flag, giving $R_{\text{gov}} \leq 1/3$. A control-plane PR can have at most two active flags, giving
$R_{\text{gov}} \leq 2/3 + \kappa$.

We set $\kappa = 0.25$. This coefficient captures the additional structural risk of changing governance mechanisms, even when the individual change appears procedurally compliant. A simple workflow modification may be inexpensive to implement but still affect many later builds or deployments. For our initial model, we leave it static, but it can be a dynamically adjusted coefficient per PR once we have more knowledge of the actual (not assumed) consequences of certain types of governance changes.

\paragraph{Layer 4: Authority-Aware Joint Assessment} 
\bigskip
This layer extracts data to calculate the metrics in the other layers. The distinction between authority scopes determines how the governance term is interpreted:

\begin{itemize}
    \item \textbf{Data-plane changes} operate within existing rules, including bug fixes, features, tests, and dependency updates.
    \item \textbf{Control-plane changes} modify the rules, including workflows, approval gates, permission settings, secrets, and deployment policies.
\end{itemize}

\section{Methodology}
\label{sec:methodology}

We evaluate the $P(M)$ framework by collecting and preparing the AIDev data, classifying PRs by authority scope and validating the classifier, computing $C_{\text{exec}}$, $R_{\text{gov}}$, and $P(M)$, and comparing the resulting scores across authority scopes and agents.

\subsection{The AIDev Dataset}
\label{sec:dataset}

We use the AIDev dataset~\cite{aidev,aidev-teammates} because it provides a large corpus of PRs authored by autonomous coding agents. The dataset is publicly available on Hugging Face\footnote{\url{https://huggingface.co/datasets/hao-li/AIDev}}, with companion code hosted on GitHub\footnote{\url{https://github.com/SAILResearch/AI_Teammates_in_SE3}}. AIDev contains 932,791 agent-authored PRs from 116,211 repositories and records interactions involving 72,189 human developers. Table~\ref{tab:dataset} reports the number of PRs and repositories associated with each agent.

\begin{table}[!ht]
\centering
\caption{Composition of the AIDev dataset by autonomous coding agent.}
\label{tab:dataset}
\begin{tabular}{lrr}
\toprule
\textbf{Agent} & \textbf{\#PRs} & \textbf{\#Repositories} \\
\midrule
OpenAI Codex & 814,522 & 84,704 \\
GitHub Copilot & 50,447 & 14,492 \\
Cursor & 32,941 & 12,699 \\
Devin & 29,744 & 4,747 \\
Claude Code & 5,137 & 1,915 \\
\midrule
\textbf{Total} & \textbf{932,791} & \textbf{116,211} \\
\bottomrule
\end{tabular}
\end{table}

For each PR, the dataset includes fields such as title, description, author, creation time, merge time, closure time, merge status, and, for some records, file-level changes. These fields support the construction of the execution-cost and governance-risk measures defined in Section~\ref{sec:pm-framework}. The agent identifier also makes it possible to compare risk profiles across implementations.

\subsection{Full-Dataset Scoring}
\label{sec:tiers}

We calculate $C_{\text{exec}}$, $R_{\text{gov}}$, and $P(M)$ for every one of the 932,791 PRs. The processing pipeline produces a sequence of intermediate artifacts: raw records, authority-scope classifications with confidence values, extracted metrics, normalized metrics, and final penalty scores.

Because file-level information is available for only a subset of the dataset, authority scope is determined primarily from the PR title and description, corresponding to the \texttt{title} and \texttt{body} fields in \texttt{all\_pull\_request}. The classifier matches this text against a curated dictionary of control-plane indicators.

A PR is classified as \textbf{data-plane} when it concerns application code, business logic, tests, documentation, or dependencies without indicating a change to CI/CD governance. Examples include bug fixes, feature implementations, test additions, dependency updates, and schema changes that do not affect approval or deployment policies. Operationally, data-plane classification is the default when no control-plane indicator is found.

A PR is classified as \textbf{control-plane} when its title or description indicates a change to workflows, permissions, secrets, deployment rules, or related governance structures. The classifier uses two categories of indicators:

\begin{itemize}
    \item \textit{Strong indicators}, where one match is sufficient, including
    \texttt{.github/workflows}, \texttt{Jenkinsfile}, \texttt{ci pipeline},
    \texttt{approval gate}, \texttt{branch protection}, \texttt{permissions:},
    \texttt{deployment policy}, \texttt{terraform}, \texttt{helm chart},
    \texttt{kubernetes manifest}, \texttt{canary}, and \texttt{blue-green}.

    \item \textit{Weak indicators}, where at least two distinct matches are required, including
    \texttt{workflow}, \texttt{pipeline}, \texttt{deploy}, \texttt{secrets},
    \texttt{rollback}, \texttt{staging}, \texttt{approval}, and
    \texttt{infrastructure}.
\end{itemize}

Classification confidence is recorded to keep track which rule produced the label. A strong-indicator match is assigned medium confidence, whereas a combination of weak indicators is assigned low confidence. These confidence values are retained in the released per-PR results.

\subsection{Classification and Validation}
\label{sec:validation}

The classification procedure consists of two steps.

\paragraph{Step 1: Automatic Classification}

The classifier scans the title and description of each PR and applies the following deterministic rules:

\begin{itemize}
    \item If at least one strong control-plane indicator or at least two distinct weak indicators are present, classify the PR as \texttt{control-plane}.
    \item Otherwise, classify the PR as \texttt{data-plane}.
\end{itemize}

Because the rules are deterministic, the classification can be reproduced directly from the dataset. 

\paragraph{Step 2: Manual Validation}

The author independently labeled a stratified sample of 34 PRs. Comparison with the automatic classification produced 94.1\% raw agreement and Cohen's $\kappa = 0.72$ towards the automated classification, conventionally interpreted as substantial agreement. 

\subsection{Metric Extraction}

After assigning authority scope, we derive the following measures from the available PR fields:

\begin{itemize}
    \item \textbf{Effort}, $\hat{C}_{\text{time}}$: an ordinal estimate derived from the length and structure of the PR description. PRs are classified as trivial (0.25), focused (0.55), or complex (0.80).

    \item \textbf{Latency}, $\hat{W}_{\text{time}}$: the interval between
    \texttt{created\_at} and either \texttt{merged\_at} or \texttt{closed\_at}. The value is log-normalized using a 30-day moving window. PRs that remain open at the time of analysis receive a value of 0.

    \item \textbf{Rejection}, $R_{\text{override}}$: set to 1 when the PR state is \texttt{closed} and \texttt{merged\_at} is absent.

    \item \textbf{Policy violation}, $R_{\text{policy}}$: set to 1 when a control-plane PR is merged in less than five minutes, or when it is merged despite failure signals. This serves as an operational proxy for a control-plane change merged without an effective authorization gate, based on heuristics from our previous studies \cite{ochodek2025acora}.

    \item \textbf{Regression}, $R_{\text{regress}}$: set to 1 when a merged PR contains failure-related evidence, such as failing checks, broken builds, or regression signals.
\end{itemize}

We use these measures directly to calculate $C_{\text{exec}}$ and $R_{\text{gov}}$.

\subsection{Illustrative Examples}
\label{sec:data-examples}

Table~\ref{tab:examples} shows two illustrative records after classification and metric extraction. The titles have been paraphrased. The first PR is a routine data-plane change; the second changes a deployment workflow. The control-plane PR has a lower execution cost than the data-plane PR (0.31 vs. 0.42), meaning it merged quickly and carried little risk. Its higher joint penalty therefore comes from the governance-risk term. This example shows why execution efficiency alone is insufficient: a small, rapidly merged change can still alter consequential governance rules.

\begin{table}[!ht]
\centering
\caption{Illustrative classified PR records: a routine data-plane fix and a control-plane pipeline change.}
\label{tab:examples}
\setlength{\tabcolsep}{4pt}
\begin{tabularx}{\linewidth}{p{3.4cm}X lccc}
\toprule
\textbf{PR title (paraphrased)} &
\textbf{Representative paths} &
\textbf{Scope} &
$C_{\text{exec}}$ &
$R_{\text{gov}}$ &
$P(M)$ \\
\midrule
Add retry logic to DB client &
\texttt{lib/db.py}, \texttt{tests/db\_test.py} &
Data &
0.42 &
0.00 &
0.17 \\
Streamline deployment approvals &
\texttt{.github/workflows/ deploy.yml} &
Control &
0.31 &
0.58 &
0.47 \\
\bottomrule
\end{tabularx}
\end{table}

The complete analysis is stratified by five agent types and two authority scopes, producing ten groups for comparison.

\section{Results}
\label{sec:results}

All results in this section are calculated from the complete set of 932,791 scored PRs.

\subsection{Descriptive Statistics}
\label{sec:descriptive}

We summarize dataset composition and control-plane prevalence by agent in Table~\ref{tab:descriptive-composition}.

\begin{table}[!htb]
\centering
\caption{Dataset composition and control-plane share by agent (all 932,791 scored PRs).}
\label{tab:descriptive-composition}
\small
\setlength{\tabcolsep}{6pt}
\begin{tabularx}{\linewidth}{lrr}
\toprule
\textbf{Measure} & \textbf{Count} & \textbf{Share (\%)} \\
\midrule
Data-plane PRs & 895,382 & 96.0 \\
Control-plane PRs & 37,409 & 4.0 \\
\midrule
\textbf{Agent} & \textbf{---} & \textbf{Control-plane share (\%)} \\
\midrule
Codex & --- & 2.1 \\
Cursor & --- & 6.2 \\
Devin & --- & 14.9 \\
Copilot & --- & 25.2 \\
Claude Code & --- & 27.1 \\
\midrule
Overall merged PRs & --- & 84.7 \\
\bottomrule
\end{tabularx}
\end{table}

As expected, control-plane activity is uncommon relative to ordinary coding work, but 4.0\% still corresponds to more than 37,000 PRs. The differences between agents suggest that control-plane activity is not an unavoidable property of agent-authored development. It varies with the agent implementation and possibly with how the agent is used. Merge rates also differ by authority scope. Control-plane PRs are merged less frequently than data-plane PRs, at 73.2\% compared with 85.2\%. This suggests that reviewers treat control-plane changes more cautiously, although most are still accepted.

Across the full dataset, the mean value of $P(M)$ is 0.229, with a median of 0.167 and a standard deviation of 0.127. Values range from 0.050 to 0.910. Most PRs (92.2\%) score below 0.4, while 1.2\% score at least 0.6. A total of 775 PRs exceed the extreme-risk threshold of 0.8.

These results indicate that most agent-authored PRs are routine, low-penalty changes. Risk is concentrated in the upper tail rather than around the typical case. At this scale, however, even a small tail contains hundreds of consequential changes, making per-PR assessment useful.

\subsection{RQ1: Authority Scope}

As shown in Table \ref{tab:scope}, the control-plane PRs have a mean $P(M)$ of 0.498, compared with 0.218 for data-plane PRs. This is a 2.3$\times$ difference. Mean governance risk is 0.530 for control-plane PRs and 0.135 for data-plane PRs, a 3.9$\times$ difference. Mean execution cost differs less substantially, by a factor of 1.3.

Policy-violation flags occur in 46.6\% of all control-plane PRs. Among merged control-plane PRs, the corresponding proportion is 63.6\%, or 17,415 of 27,383 PRs. All 775 PRs with $P(M) > 0.8$ are control-plane changes. No data-plane PR reaches the extreme-risk range.

\begin{table}[!htb]
\centering
\caption{$P(M)$ components by authority scope for all 932,791 PRs. $R_{\text{policy}}$ applies only to merged control-plane PRs by construction.}
\label{tab:scope}
\small
\setlength{\tabcolsep}{4pt}
\begin{tabular}{lrrrrr}
\toprule
\textbf{Scope} &
\textbf{n} &
\textbf{$P(M)$} &
\textbf{$R_{\text{gov}}$} &
\textbf{$C_{\text{exec}}$} &
\textbf{Policy viol.} \\
\midrule
Data-plane & 895,382 & 0.218 & 0.135 & 0.343 & --- \\
Control-plane & 37,409 & 0.498 & 0.530 & 0.449 & 46.6\% \\
\midrule
\textbf{Ratio} & & $2.3\times$ & $3.9\times$ & $1.3\times$ & \\
\bottomrule
\end{tabular}
\end{table}

\textbf{Interpretation.}
Authority scope separates the risk distributions more clearly than execution cost. The structural coefficient $\kappa = 0.25$ can explain no more than 0.25 of the observed 0.395 difference in mean $R_{\text{gov}}$. The remaining difference results from policy and regression indicators rather than from the control-plane premium alone.

The concentration of all extreme-risk PRs in the control-plane category is particularly notable. Under the proposed model, control-plane authority is a necessary condition for reaching the highest-risk range.

\begin{quote}
\textbf{Answer to RQ1.}
Control-plane PRs have systematically higher penalties than data-plane PRs. Their mean $P(M)$ is 2.3$\times$ higher, their mean governance risk is 3.9$\times$ higher, and they account for all 775 PRs in the extreme-risk category. Policy-violation flags occur in 46.6\% of control-plane PRs and in 63.6\% of merged control-plane PRs.
\end{quote}

\subsection{RQ2: Variation Across Agents}

Mean $P(M)$ differs by a factor of 2.0 across agents, ranging from 0.166 for Cursor to 0.328 for Copilot. Mean $R_{\text{gov}}$ differs by a factor of 3.2, from 0.059 to 0.187, as shown in Table~\ref{tab:agent-pm}.

The agents form two broad execution-cost groups. Cursor and Codex have mean $C_{\text{exec}}$ values close to 0.33, whereas Claude Code, Devin, and Copilot have values between 0.51 and 0.55. Despite having almost identical mean execution costs, Cursor and Codex differ by a factor of 2.6 in mean governance risk, at 0.059 and 0.152, respectively.

The execution-cost distributions show a similar division. Cursor and Codex are concentrated around 0.27--0.30, corresponding to trivial tasks and short resolution times. Claude Code, Devin, and Copilot are distributed more broadly between 0.4 and 0.85. Devin also has additional mass in the complex-task range. Their $P(M)$ distributions reflect these differences: Cursor is concentrated at lower values, Codex is bimodal, and Claude Code, Devin, and Copilot have broader distributions with longer upper tails.

\begin{table}[!htb]
\centering
\caption{$P(M)$ framework scores by autonomous coding agent, sorted by mean $P(M)$.}
\label{tab:agent-pm}
\small
\setlength{\tabcolsep}{4pt}
\begin{tabular}{lrrrrr}
\toprule
\textbf{Agent} &
\textbf{PRs (n)} &
\textbf{Merge \%} &
\textbf{$C_{\text{exec}}$} &
\textbf{$R_{\text{gov}}$} &
\textbf{$P(M)$} \\
\midrule
Cursor & 32,941 & 70.5 & 0.326 & 0.059 & 0.166 \\
OpenAI Codex & 814,522 & 87.7 & 0.327 & 0.152 & 0.222 \\
Claude Code & 5,137 & 76.8 & 0.512 & 0.187 & 0.317 \\
Devin & 29,744 & 64.4 & 0.536 & 0.171 & 0.317 \\
Copilot & 50,447 & 59.1 & 0.549 & 0.181 & 0.328 \\
\midrule
\textbf{Spread} & & & & $3.2\times$ & $2.0\times$ \\
\bottomrule
\end{tabular}
\end{table}

\textbf{Interpretation.}
The merge rate does not reliably approximate governance risk. Codex has the highest merge rate, 87.7\%, but a mid-range governance-risk score. Copilot has the lowest merge rate, 59.1\%, and the highest mean joint penalty. Execution cost may reflect differences in task selection, such as small, focused patches versus larger or more complex changes. The contrast between Cursor and Codex is especially informative: the two agents have similar mean execution costs but substantially different governance-risk profiles. These differences suggest that a single authority policy applied uniformly to every agent is unlikely to be well calibrated.

\begin{quote}
\textbf{Answer to RQ2.}
Governance risk differs across agent implementations. Mean $P(M)$ varies by a factor of 2.0, from 0.166 for Cursor to 0.328 for Copilot, while mean $R_{\text{gov}}$ varies by a factor of 3.2. These differences are not explained by execution cost or merge rate, indicating that authority policies should be calibrated at the agent level.
\end{quote}

\subsection{PR-Level Ranking and the Relationship Between Cost and Risk}

\begin{figure}[!htb]
\centering
\begin{minipage}{0.48\textwidth}
\centering
\includegraphics[width=\textwidth]{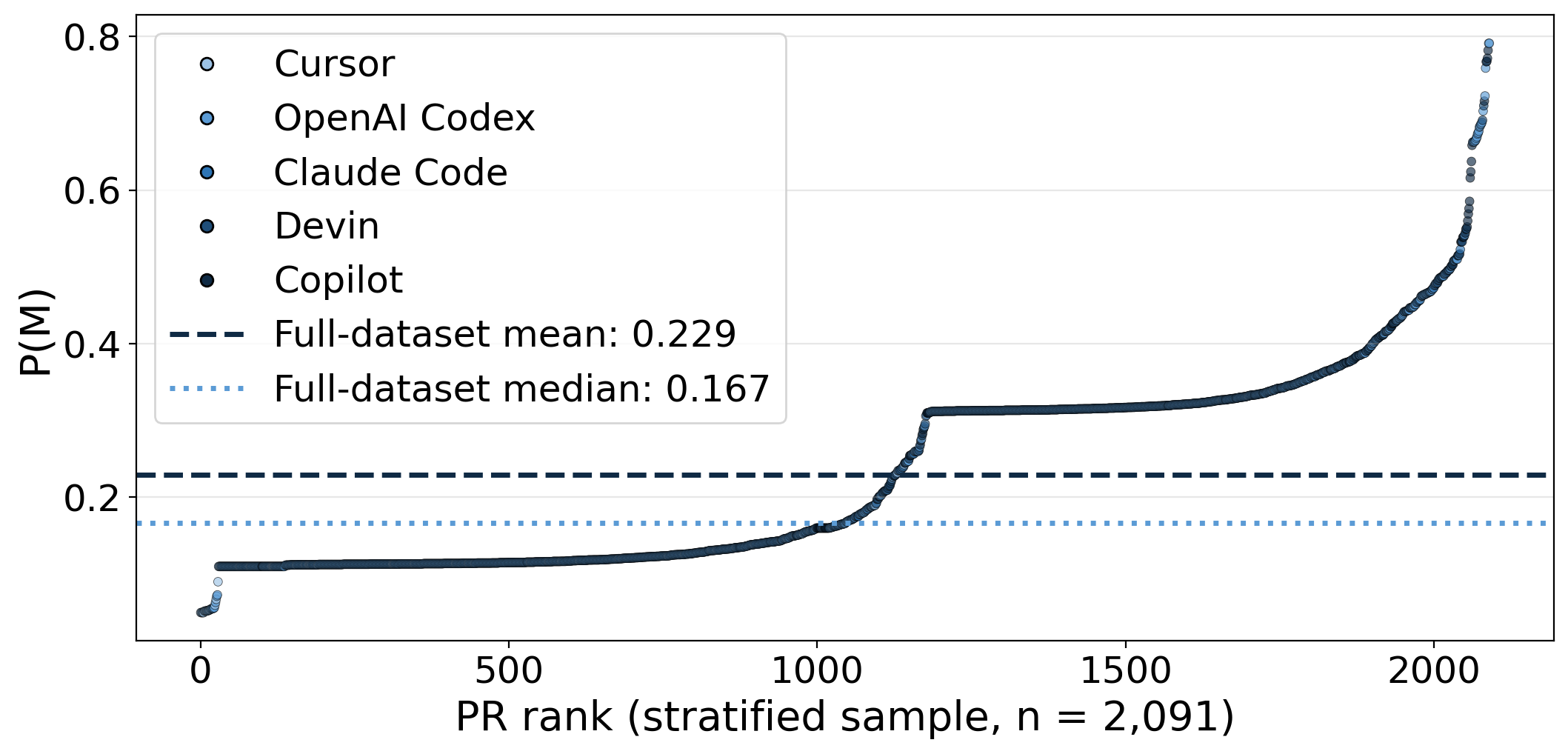}
\end{minipage}\hfill
\begin{minipage}{0.48\textwidth}
\centering
\includegraphics[width=\textwidth]{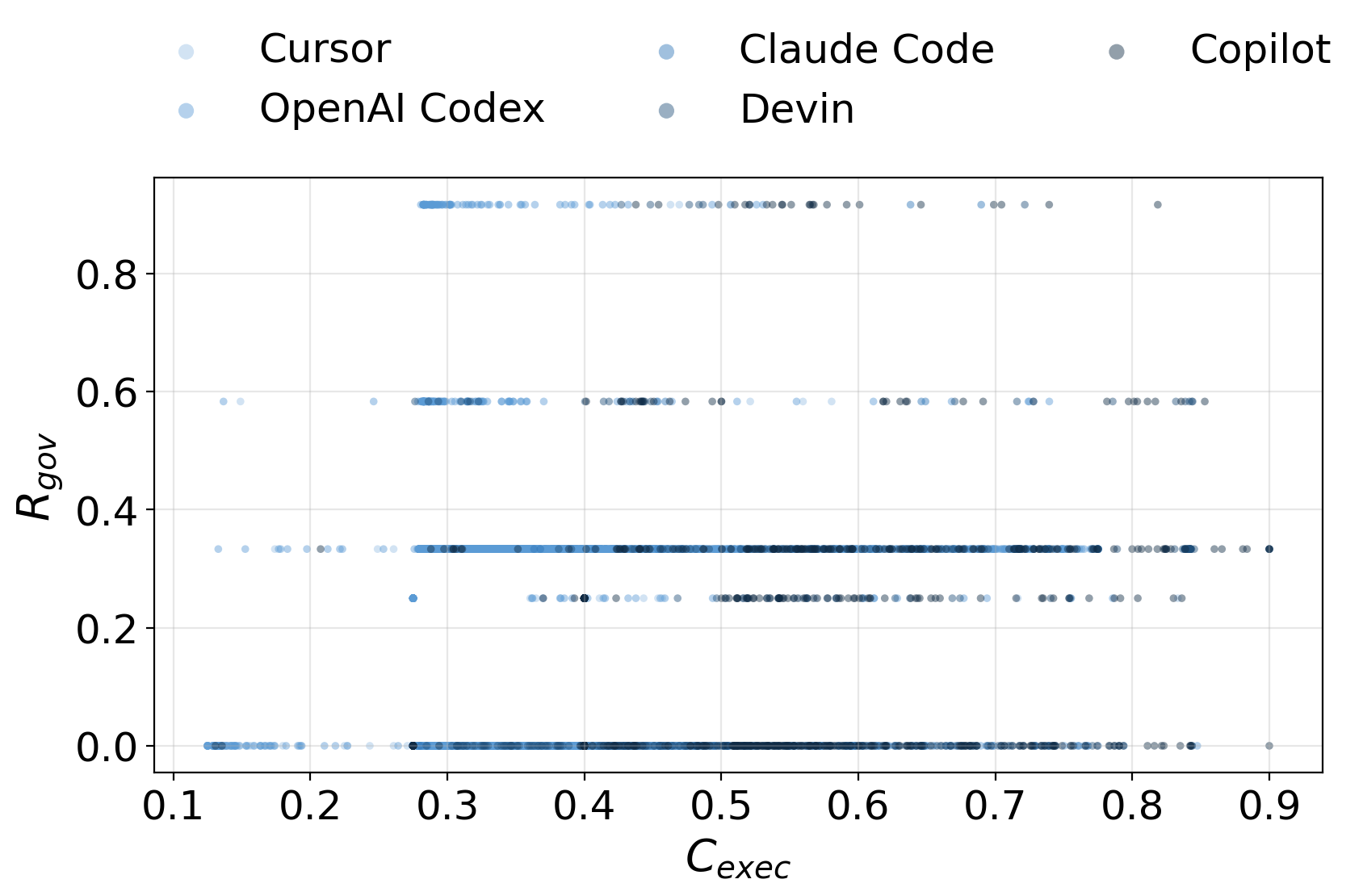}
\end{minipage}
\caption{Left: PRs ranked by $P(M)$ in a stratified random sample ($n = 2{,}091$), with full-dataset mean and median shown as dashed lines. Right: execution cost and governance risk for a random sample of 12,000 PRs, grouped by agent. The horizontal bands arise from combinations of the binary governance indicators and the $\kappa$ control-plane premium.}
\label{fig:combined-rank-scatter}
\end{figure}

The left-hand side of Figure~\ref{fig:combined-rank-scatter} ranks a stratified random sample of 2,091 PRs by $P(M)$. The sample is proportional to the agent distribution and includes at least ten PRs from each agent. The ranking contains a long lower plateau near $P(M) \approx 0.11$, followed by an increase through the full-dataset mean, a second concentration near 0.31, and a steep rise in the final decile toward 0.8. PRs from different agents are interleaved across the ranking.

The right-hand side plots $C_{\text{exec}}$ against $R_{\text{gov}}$ for a random sample of 12,000 PRs. Governance risk appears in discrete bands at approximately 0, 0.25, 0.33, 0.58, and 0.92. The Pearson correlation between execution cost and governance risk in the full dataset is $r = 0.17$.

\textbf{Interpretation.}
The interleaving of agents in the ranking indicates that $P(M)$ is primarily a measure of individual changes rather than an agent ranking. Each agent contributes PRs across several risk levels. The upper tail is driven more strongly by authority scope than by agent identity.

The step-like shape of the ranking follows from the discrete governance indicators. Each additional flag, or the addition of the control-plane premium, moves a PR to a higher risk band.

The correlation of $r = 0.17$ implies that execution cost and governance risk share only a small proportion of their variance. Cost and latency therefore cannot serve as reliable proxies for governance. In particular, inexpensive and quickly resolved PRs may still carry substantial governance risk.

\subsection{Distribution of Governance Risk}
\begin{wrapfigure}{r}{0.5\textwidth}
    \centering
    \includegraphics[width=\linewidth]{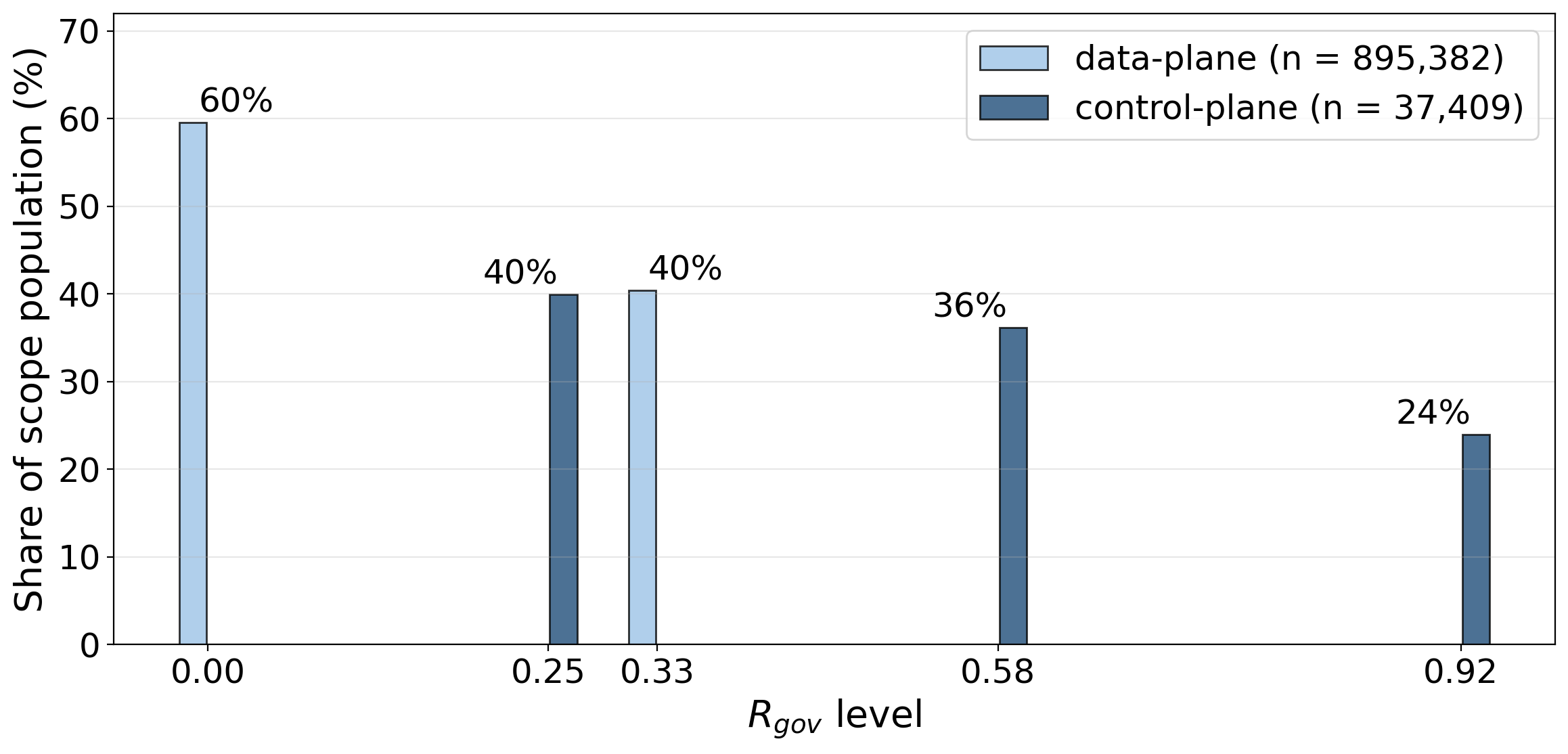}
    \caption{Distribution of governance risk $R_{\text{gov}}$ by authority scope. The percentages are rounded to full numbers.}
    \label{fig:rgov-distribution}
\end{wrapfigure}

Figure \ref{fig:rgov-distribution} shows that within each authority scope, $R_{\text{gov}}$ takes a limited set of values. Among data-plane PRs, 59.6\% have $R_{\text{gov}} = 0$, as they do not modify the pipeline. The remaining 40.4\% have $R_{\text{gov}} = 0.33$, indicating a potential pipeline modification (but one flag only). By construction, data-plane governance risk cannot exceed 0.33.

Control-plane PRs are distributed across three main values:

\begin{itemize}
    \item 39.9\% have $R_{\text{gov}} = 0.25$, corresponding to the structural premium alone: the fact that we found trace/flag of changing the control plane, but could not quantify its impact/risk;
    \item 36.1\% have $R_{\text{gov}} = 0.58$, corresponding to the premium and one additional flag;
    \item 24.0\% have $R_{\text{gov}} = 0.92$, corresponding to the premium and two additional flags.
\end{itemize}

\textbf{Interpretation.}
The discrete values distinguish between frequent control-plane activity and problematic control-plane activity. A score of 0.25 identifies a control-plane modification without an additional observed violation, whereas values of 0.58 and 0.92 indicate one or two further governance concerns.







\subsection{Threats to Validity}

We structure the discussion according to Wohlin et al.~\cite{wohlin2012experimentation} and Staron~\cite{staron2019action}.

\textbf{Construct validity.}
The proposed measure operationalizes governance risk through a compound metric whose weights and coefficients have not yet been validated in industrial settings. The selected values affect score magnitude and should be calibrated to specific organizational contexts. In addition, rejection, rapid merge, and failure-related language are proxies for governance outcomes rather than direct observations of review quality or policy enforcement.

\textbf{Internal validity.}
We infer authority scope and task effort using heuristics based on our previous studies. Although the manual audit produced substantial agreement, classification errors may remain, particularly where PR descriptions are incomplete or do not accurately reflect the code changes. Using title and description text is also a limitation because file-level information is unavailable for much of the dataset. The selected thresholds, including the five-minute review floor and $\kappa = 0.25$, may affect estimated effect sizes.

\textbf{External validity.}
The study uses a single public dataset dominated by open-source repositories and OpenAI Codex PRs. The findings may not generalize directly to proprietary repositories, regulated development environments, or organizations with different review and deployment practices. The coefficients and thresholds should therefore be recalibrated before operational use.

\textbf{Conclusion validity.}
The analysis covers the full dataset rather than a statistical sample, reducing sampling uncertainty within AIDev. However, the lack of a comparable independent dataset limits replication and cross-dataset comparison. Future work should reproduce the analysis using other collections of agent-authored PRs and test the conclusions' sensitivity to alternative metric definitions.

\section{Conclusions}
\label{sec:conclusion}

Autonomous coding agents are increasingly involved in activities that extend beyond code generation. In some cases, they modify the CI/CD infrastructure that governs review, approval, and deployment. Such changes may address an immediate engineering problem while also altering the constraints applied to future work.

This paper introduced $P(M)$, a joint penalty model that combines execution cost and governance risk. We applied the model to all 932,791 PRs in the AIDev dataset, classified each PR by authority scope, and compared the resulting scores across five coding agents. The findings show that control-plane PRs have substantially higher governance risk and joint penalty than data-plane PRs. They account for the entire extreme-risk tail of the dataset. Governance-risk profiles also differ considerably across agent implementations and are not explained by merge rate or execution cost.


Future work will focus on three areas: evaluating the framework in industrial environments, improving authority-scope classification using file-level and diff-level evidence, and studying how agent-specific governance policies affect software-development outcomes.



\bibliographystyle{splncs04}
\bibliography{research_paper_with_results}

@book{wohlin2012experimentation,
  title={Experimentation in software engineering},
  author={Wohlin, Claes and Runeson, Per and H{\"o}st, Martin and Ohlsson, Magnus C and Regnell, Bj{\"o}rn and Wessl{\'e}n, Anders and others},
  volume={236},
  year={2012},
  publisher={Springer}
}

@inproceedings{assistance-to-agency,
  title={From Assistance to Agency: Rethinking Autonomy and Control in CI/CD Pipelines},
  author={Barnes, Marcus Emmanuel and Ghaleb, Taher A and Hassan, Safwat},
  booktitle={AIWare},
  year={2026},
  notes={arXiv preprint arXiv:2605.07062}
}

@inproceedings{cicd-stages,
  title={“Good” and “Bad” Failures in Industrial CI/CD--Balancing Cost and Quality Assurance},
  author={Sun, Simin and Friberg, David and Staron, Miroslaw},
  booktitle={Euromicro Conference on Software Engineering and Advanced Applications},
  pages={75--84},
  year={2025},
  organization={Springer}
}

@article{mape-k,
  author  = {Kephart, Jeffrey O. and Chess, David M.},
  title   = {The Vision of Autonomic Computing},
  journal = {Computer},
  volume  = {36},
  number  = {1},
  pages   = {41--50},
  year    = {2003}
}

@article{aidev-teammates,
  title={The rise of ai teammates in software engineering ({SE}) 3.0: How autonomous coding agents are reshaping software engineering},
  author={Li, Hao and Zhang, Haoxiang and Hassan, Ahmed E},
  journal={arXiv preprint arXiv:2507.15003},
  year={2025}
}

@article{aidev,
  author        = {Li, Hao and Zhang, Haoxiang and Hassan, Ahmed E.},
  title         = {{AIDev}: Studying {AI} Coding Agents on {GitHub}},
  year          = {2026},
  eprint        = {2602.09185},
  archivePrefix = {arXiv},
  primaryClass  = {cs.SE},
  url           = {https://arxiv.org/abs/2602.09185},
  note          = {Dataset: \url{https://huggingface.co/datasets/hao-li/AIDev}}
}

@article{agent-readmes,
  title={Agent {READMEs}: An Empirical Study of Context Files for Agentic Coding},
  author={Chatlatanagulchai, Worawalan and Li, Hao and Kashiwa, Yutaro and Reid, Brittany and Thonglek, Kundjanasith and Leelaprute, Pattara and Rungsawang, Arnon and Manaskasemsak, Bundit and Adams, Bram and Hassan, Ahmed E and others},
  journal={arXiv preprint arXiv:2511.12884},
  year={2025}
}

@article{agentic-refactoring,
  title={Agentic Refactoring: An Empirical Study of {AI} Coding Agents},
  author={Horikawa, Kosei and Li, Hao and Kashiwa, Yutaro and Adams, Bram and Iida, Hajimu and Hassan, Ahmed E},
  journal={arXiv preprint arXiv:2511.04824},
  year={2025}
}

@article{message-consistency,
  title={Analyzing message-code inconsistency in ai coding agent-authored pull requests},
  author={Gong, Jingzhi and Pinna, Giovanni and Bian, Yixin and Zhang, Jie M},
  booktitle={Proceedings of the 23rd International Conference on Mining Software Repositories},
  pages={827--831},
  year={2026}
}

@article{test-code,
  title={Do Autonomous Agents Contribute Test Code? A Study of Tests in Agentic Pull Requests},
  author={Haque, Sabrina and Ingale, Sarvesh and Csallner, Christoph},
  journal={arXiv preprint arXiv:2601.03556},
  year={2026}
}

@article{review-effort,
  title={Early-stage prediction of review effort in ai-generated pull requests},
  author={Minh, Dao Sy Duy and Kiet, Huynh Trung and Quy, Nguyen Lam Phu and Hoa, Pham Phu and Nguyen, Tran Chi and Duong, Nguyen Dinh Ha and Tran, Truong Bao},
  booktitle={Proceedings of the 23rd International Conference on Mining Software Repositories},
  pages={954--958},
  year={2026}
}

@article{security-agentic,
  title={Security in the Age of {AI} Teammates: An Empirical Study of Agentic Pull Requests on GitHub},
  author={Siddiq, Mohammed Latif and Zhao, Xinye and Lopes, Vinicius Carvalho and Casey, Beatrice and Santos, Joanna},
  journal={arXiv preprint arXiv:2601.00477},
  year={2026}
}

@article{code-optimization,
  title={How do agents perform code optimization? An empirical study},
  author={Peng, Huiyun and Qiu, Antonio Zhong and M{\'e}ndez, Ricardo Andr{\'e}s Calvo and Kalu, Kelechi G and Davis, James C},
  booktitle={Proceedings of the 23rd International Conference on Mining Software Repositories},
  pages={732--736},
  year={2026}
}

@article{security-risks-deps,
  title={Understanding Security Risks of {AI} Agents' Dependency Updates},
  author={Singla, Tanmay and {\c{C}}akar, Berk and Amusuo, Paschal C and Davis, James C},
  journal={arXiv preprint arXiv:2601.00205},
  year={2026}
}

@article{will-it-survive,
  title={Will It Survive? Deciphering the Fate of {AI-Generated} Code in Open Source},
  author={Rahman, Musfiqur and Shihab, Emad},
  journal={arXiv preprint arXiv:2601.16809},
  year={2026}
}

@article{library-usage,
  title={A Study of Library Usage in Agent-Authored Pull Requests},
  author={Twist, Lukas and Zhang, Jie M},
  booktitle={Proceedings of the 23rd International Conference on Mining Software Repositories},
  pages={722--726},
  year={2026}
}

@article{energy-concerns,
  title={How do agentic ai systems deal with software energy concerns? a pull request-based study},
  author={Mitul, Tanjum Motin and Mazumder, Md Masud and Opu, Md Nahidul Islam and Chowdhury, Shaiful},
  booktitle={Proceedings of the 23rd International Conference on Mining Software Repositories},
  pages={883--887},
  year={2026}
}

@incollection{staron2019action,
  title={Action research as research methodology in software engineering},
  author={Staron, Miroslaw},
  booktitle={Action Research in Software Engineering: Theory and Applications},
  pages={15--36},
  year={2019},
  publisher={Springer}
}

@article{issue-ready,
  title={What Makes a {GitHub} Issue Ready for {Copilot}?},
  author={Sayagh, Mohammed},
  journal={arXiv preprint arXiv:2512.21426},
  year={2025}
}

@article{review-themes,
  title={Understanding dominant themes in reviewing agentic ai-authored code},
  author={Haider, Md Asif and Zimmermann, Thomas},
  booktitle={Proceedings of the 23rd International Conference on Mining Software Repositories},
  pages={742--746},
  year={2026}
}

@article{ochodek2025acora,
  title={Acora--a platform for automating code review tasks},
  author={Ochodek, Miroslaw and Staron, Miroslaw},
  journal={e-Informatica Software Engineering Journal},
  volume={19},
  number={1},
  year={2025}
}

\end{document}